\documentclass[12pt]{article}
\usepackage{graphicx,latexsym}

\def\a{\alpha} \def\b{\beta} \def\dl{\delta} \def\s{\sigma}     \def\lam{\lambda}  \def\gm{\gamma} \def\Gm{\Gamma} \def\om{\omega} \def\Om{\Omega} \def\nb{\nabla} \def\sq{\sqrt} 
 \def\pd{\partial} \def\pp{\prime} \def\half{\frac{1}{2}} \def\fr{\frac}

   \def\V3{{\rm V}_3}

 \def\bg{{\bar g}} \def\bDelta{{\bar \Delta}} 
\def\bE{{\bar E}}

\def\lap3{~| \!\!\! \partial^2} \def\dlap3{~| \!\!\! \partial^4}

\def\lang{\langle} \def\rang{\rangle}
\def\barb{{\bar \beta}}

\begin{document}

\begin{titlepage}

\begin{flushright}
March 2014
\end{flushright}

\vspace{5mm}

\begin{center}
{\Large {\bf Determination of Gravitational Counterterms Near Four Dimensions from RG Equations}}
\end{center}

\vspace{5mm}

\begin{center}
{\sc Ken-ji Hamada}\footnote{E-mail address: hamada@post.kek.jp; URL: http://research.kek.jp/people/hamada/}
\end{center}

\begin{center}
{\it Institute of Particle and Nuclear Studies, KEK, Tsukuba 305-0801, Japan} \\ and \\
{\it Department of Particle and Nuclear Physics, The Graduate University for Advanced Studies (Sokendai), Tsukuba 305-0801, Japan}
\end{center}

\begin{abstract}
The finiteness condition of renormalization gives a restriction on the form of the gravitational action. By reconsidering the Hathrell's RG equations for massless QED in curved space, we determine the gravitational counterterms and the conformal anomalies as well near four dimensions. As conjectured for conformal couplings in 1970s, we show that at all orders of the perturbation they can be combined into two forms only: the square of the Weyl tensor in $D$ dimensions and
\begin{eqnarray*}
   E_D=G_4 +(D-4)\chi(D)H^2 -4\chi(D) \nabla^2 H , 
\end{eqnarray*} 
where $G_4$ is the usual Euler density, $H=R/(D-1)$ is the rescaled scalar curvature and $\chi(D)$ is a finite function of $D$ only. The number of the dimensionless gravitational couplings is also reduced to two. $\chi(D)$ can be determined order by order in series of $D-4$, whose first several coefficients are calculated. It has a universal value of $1/2$ at $D=4$. The familiar ambiguous $\nabla^2 R$ term is fixed. At the $D \to 4$ limit, the conformal anomaly $E_D$ just yields the combination $E_4=G_4-2\nabla^2 R/3$, which induces Riegert's effective action.  
\end{abstract}

\vspace{5mm}

\end{titlepage}

\section{Introduction}
\setcounter{equation}{0}
\noindent

Recently, it has become increasingly important to understand how to include gravity within the framework of quantum field theory, especially when we consider models of the early universe such as inflation and quantum gravity. How to handle gravitational divergences is one of the most significant problems in this area.

We here consider gravitational counterterms for a four-dimensional quantum field theory in curved spacetime. Usually, we consider three independent gravitational counterterms and associated three dimensionless coupling constants. For a conformally coupled theory, however, there is an old conjecture in 1970s that gravitational divergences are simply renormalized by using conformally invariant counterterms: the square of the Weyl tensor and the Euler density \cite{cd, ddi, duff, dh, ds}.

At a later time, however, the seemingly negative result that the $R^2$-divergence appears in calculations of 3-loop or more using dimensional regularization was reported by Brown and Collins \cite{bc} and then by Hathrell \cite{hathrell-scalar,hathrell-QED} and Freeman \cite{freeman}. On the other hand, Hathrell also showed in his paper that two counterterms of the Euler density and $R^2$ are related to each other through renormalization group (RG) equations.

In this paper, we revive these old works and reconsider the meaning of their RG equations. It is then revealed that the appearance of the $R^2$ divergence is simply a dimensional artifact coming from the fact that there is an indefiniteness in $D$-dimensional gravitational counterterms that reduce to conformally invariant ones at four dimensions. We see that the Euler density and $R^2$ counterterms can be unified and the number of gravitational couplings can be reduced for conformally coupled theories. At the same time, the ambiguous term $\nb^2 R$ in conformal anomaly can be fixed completely.

In four dimensions, the conformal anomaly is obtained by regularizing the divergent quantity $\dl^4(0)=\lang x| x^\pp \rang |_{x^\pp \to x}$ coming from the path integral measure. On the other hand, if we use dimensional regularization, the result is independent of how to choose the measure because of $\dl^D(0)=\int d^D k=0$. This fact suggests that in dimensional regularization the information of the measure is contained between $D$ and $4$ dimensions. Thus, it is significant to determine the $D$-dependence of the counterterms. It is one of the aims of this study as well.

\section{QED in Curved Space}
\setcounter{equation}{0}
\noindent

As a prototype of conformally coupled quantum field theory, we here consider massless QED in curved space, because it is the simplest theory with unambiguously fixed coupling between fields and gravity.

To begin with, we define the theory using dimensional regularization and summarize the notation and conventions. The action of QED in curved space is defined by
\begin{eqnarray*}
    S = \int d^D x \sq{g} \biggl\{  \fr{1}{4}F_{0\mu\nu}F_0^{\mu\nu} + i{\bar \psi}_0 D\!\!\!\!/ \psi_0 
          + \fr{1}{2\xi_0} \left( \nb^\mu A_{0\mu} \right)^2  
          + a_0 F_D + b_0 G_4 + c_0 H^2 \biggr\} ,
\end{eqnarray*}
where we consider the Wick-rotated Euclidean space. The quantity with the subscript $0$ denotes the bare quantity before renormalization. The Dirac operator is defined by $D\!\!\!\!/=e^{\mu a}\gm_{a}D_{\mu}$, where $e_{\mu}^{~a}$ are vierbein fields in $D$ dimensions satisfying $e_{\mu}^{~a}e_{\nu a}=g_{\mu\nu}$ and $e_{\mu a}e^{\mu}_{~b}=\dl_{ab}$. The Dirac matrices are normalized as $\{ \gm_{a}, \gm_{b} \}=-2\dl_{ab}$. The covariant derivative acting fermions is defined by $D_{\mu}=\pd_{\mu}+\half \om_{\mu ab}\Sigma^{ab}+ie_0 A_{0\mu}$, where the connection 1-form and Lorentz generators are given by $\om_{\mu ab}=e^{\nu}_{~a}(\pd_{\mu}e_{\nu b}-\Gamma^{\lam}_{~\mu\nu}e_{\lam b})$ and $\Sigma^{ab}=-\fr{1}{4}[\gm^{a},\gm^{b}]$, respectively. The ghost action is disregarded here because we discuss coupling-dependent parts only.

For the moment, we consider the three types of gravitational counterterms adopted by Hathrell in his original paper \cite{hathrell-QED}. The term $F_D$ is the square of the Weyl tensor in $D$ dimensions defined by 
\begin{eqnarray}
    F_D = R_{\mu\nu\lam\s}R^{\mu\nu\lam\s} -\fr{4}{D-2}R_{\mu\nu}R^{\mu\nu}
                         +\fr{2}{(D-1)(D-2)}R^2 .
         \label{square of Weyl tensor in D dimensions}
\end{eqnarray}
The term $G_4$ is the Euler density and $H$ is the scalar curvature scaled by a $D$-dependent factor, respectively, as
\begin{eqnarray*}
    G_4=R_{\mu\nu\lam\s}^2-4R_{\mu\nu}^2+R^2, \qquad H=\fr{R}{D-1} .
\end{eqnarray*}
Our sign convention of $a_0$, $b_0$ and $c_0$ is different from \cite{hathrell-QED}. In the later sections, we will show that the last two counterterms can be combined into one at last by relating $b_0$ and $c_0$. On the other hand, $a_0$ does not mix with the others.

The renormalization factors for quantum fields are defined by
\begin{eqnarray*}
    A_{0\mu} = Z_3^{1/2}A_{\mu}, \qquad   \psi_0 = Z_2^{1/2}\psi 
\end{eqnarray*}
and the renormalizations of the coupling constant and the gauge-fixing parameter are defined by
\begin{eqnarray*}
       e_0 = \mu^{2-D/2} Z_3^{-1/2} e, \qquad \xi_0 = Z_3 \xi .
\end{eqnarray*}
Here, $\mu$ is an arbitrary mass scale to make up the loss of mass dimensions and thus the renormalized coupling $e$ is dimensionless. In the following, we mainly use the fine structure constant defined by $\a=e^2/4\pi$.

The RG equations are derived from the fact that bare quantities are independent of the arbitrary mass scale $\mu$ such as
\begin{eqnarray*}
   \mu \fr{d}{d\mu} \left({\rm bare}\right) = 0, \qquad 
     \mu \fr{d}{d\mu} = \mu \fr{\pd}{\pd \mu} + \mu \fr{d \a}{d \mu} \fr{\pd}{\pd \a}
                        + \mu \fr{d \xi}{d \mu} \fr{\pd}{\pd \xi} + \cdots .
\end{eqnarray*}
First, we consider the following equation: 
\begin{eqnarray*}
   \mu \fr{d }{d\mu} \left( \fr{e_0^2}{4\pi} \right) = 0= \fr{\mu^{4-D}}{Z_3} \a 
        \left( 4-D - \mu \fr{d}{d\mu} \log Z_3 + \fr{\mu}{\a} \fr{d\a}{d\mu} \right) .
\end{eqnarray*}
From this, the beta function for $\a$ is defined as
\begin{eqnarray*}
    \b(\a,D) \equiv \fr{1}{\a} \mu \fr{d\a}{d\mu} = (D-4) + \barb(\a) ,
\end{eqnarray*}
where $\barb= \mu d (\log Z_3)/d\mu$. If we expand the renormalization factor as $\log Z_3=\sum_{n=1}^\infty f_n(\a)/(D-4)^n$, $\barb$ is determined to be $\a \pd f_1/\pd \a$ and the equation $\pd f_{n+1}/\pd \a + \barb \pd f_n/\pd \a =0$ must be satisfied in order that the beta function is finite when the $D \to 4$ limit is taken.

In the following, we must be aware of the difference between $\b$ and $\barb$, because $1/\barb$ is finite, while
\begin{eqnarray}
   \fr{1}{\b} = \fr{1}{D-4} \left( 1 + \sum_{n=1}^\infty \fr{(-\barb)^n}{(D-4)^n} \right)
     \label{pole expansion of 1/beta}
\end{eqnarray} 
has poles in the expansion for $\a$.

The gravitational counterterms are defined by 
\begin{eqnarray*}
    a_0 &=& \mu^{D-4} \left( a + L_a \right) , \qquad L_a = \sum_{n=1}^\infty \fr{a_n(\a)}{(D-4)^n} , 
       \nonumber \\
    b_0 &=& \mu^{D-4} \left( b + L_b \right) , \qquad L_b = \sum_{n=1}^\infty \fr{b_n(\a)}{(D-4)^n} , 
       \nonumber \\
    c_0 &=& \mu^{D-4} \left( c + L_c \right) , \qquad L_c = \sum_{n=1}^\infty \fr{c_n(\a)}{(D-4)^n} , 
\end{eqnarray*}
where $L_{a,b,c}$ are the pure-pole terms whose residues are the functions of $\a$ only and $a$, $b$, $c$ are the gravitational coupling constants. The beta functions for them are defined by 
\begin{eqnarray*}
    \b_a(\a,D) \equiv  \mu \fr{d a}{d\mu}= -(D-4)a + \barb_a(\a) 
\end{eqnarray*}
and similar expressions for $b$ and $c$.

As in the case of $\b$, from the conditions that the bare coupling $a_0$ is independent of $\mu$ and $\b_a$ is finite, we obtain the expression $\barb_a(\a) = - \pd ( \a a_1)/\pd \a$ and the equation
\begin{eqnarray}
    \fr{\pd}{\pd \a} \left( \a a_{n+1} \right) + \barb \a \fr{\pd a_n}{\pd \a} = 0 
       \label{pole equation of gravity}
\end{eqnarray}
for $n \geq 1$. The similar equations also satisfy for $b_n$ and $c_n$.

When we discuss the finiteness of the theory, various normal products, namely finite composite operators, are significantly used. The normal product of dimension $4$ is constructed as a linear combination of all available composite operators of dimension less than or equal $4$ with appropriate symmetry and have to reduce to the bare field in the vanishing coupling limit. For example, $[F_{\mu\nu}F^{\mu\nu}] = ( 1 + \sum {\rm poles}) F_{0\mu\nu}F_0^{\mu\nu} + (\sum {\rm poles}) (\hbox{other operators})$, where the notation $[~]$ denotes the normal product. The derivation of this normal product is briefly summarized in Appendix A.

The trace of the energy-momentum tensor denoted by $\theta$, which is intrinsically in a bare quantity obtained by applying $\dl/\dl \Om= (2/\sq{g}) g_{\mu\nu} \dl/\dl g_{\mu\nu}$ to the action, can be written in a finite expression using the normal products as 
\begin{eqnarray}
  \theta 
     &=& \fr{D-4}{4} F_{0\mu\nu} F_0^{\mu\nu} + \fr{D-1}{2} E_{0\psi} 
         + (D-4) \left( a_0 F_D + b_0 G_4 + c_0 H^2 \right) -4c_0 \nb^2 H
                   \nonumber \\
     &=& \fr{\b}{4} \left[ F_{\mu\nu}F^{\mu\nu} \right] + \half \left( D-1 + {\bar \gm}_2 \right) [E_\psi] 
                    \nonumber \\
     &&    - \mu^{D-4} \left( \b_a F_D + \b_b G_4 + \b_c H^2 \right) 
                 -4 \mu^{D-4} (c-\s ) \nb^2 H ,
         \label{expression of Hathrell conformal anomaly}
\end{eqnarray}
where ${\bar \gm}_2=\gm_2-(D-4)\xi \pd (\log Z_2)/\pd \xi$ is a combination that becomes independent of $\xi$ and $\gm_2=\mu d (\log Z_2)/d\mu$ is the usual anomalous dimension. The normal product $[E_\psi]$ is the equation-of-motion operator for fermions defined in Appendix B. From the finiteness of the energy-momentum tensor, the $\a$-dependent function $\s$ in the last term is determined to be $\barb_c + \s \barb =0$ and $L_\s$ in (\ref{normal product of F^2}) becomes equal to $L_c$. This is the expression of the conformal anomaly derived by Hathrell.

This expression, however, has the following undesirable structure. Taking the $D \to 4$ limit, we can see that the dependence on the unspecified parameters $\mu$, $a$ and $b$ in (\ref{expression of Hathrell conformal anomaly}) disappears, but $c$ in the last term remains with a finite effect, which is known as the ambiguous $\nb^2 R$ term in the conformal anomaly. One of the aims of this paper is to remove such an ambiguity and express the conformal anomaly in a simpler form.

\section{Hathrell's RG Equations}
\setcounter{equation}{0}
\noindent

In this section, we briefly review the Hathrell's RG equations \cite{hathrell-QED}, which are derived on the basis of the RG analysis by Brown and Collins \cite{bc} combined with a study of renormalized composite operators to deduce relationship between various quantities in the theory.

\subsection{Two-point functions}
\noindent

We first consider the two-point function of the energy-momentum tensor modified as 
\begin{eqnarray*}
   {\bar \theta} = \theta - \half (D-1) [E_\psi] .
\end{eqnarray*}
Since one-point functions are dimensionally regularized to zero for a massless theory in flat space, $\lang [E_\psi(x)] P(y) \rang_{\rm flat} = \lang \dl P(y)/\dl \chi(x) \rang_{\rm flat} = 0$ is satisfied for a polynomial composite $P(y)$ in the fields ${\bar \psi}(y)$ and $\psi(y)$, where the functional derivative $\dl/\dl \chi$ is defined in Appendix B. Using this fact and the condition of the two-point function of $\theta$ (\ref{finiteness condition for theta-theta}) given in Appendix C, we obtain the following condition:
\begin{eqnarray}
     \lang {\bar \theta}(p) {\bar \theta}(-p) \rang_{\rm flat} - 8p^4 \mu^{D-4} L_c = {\rm finite} 
       \label{theta-theta determining L_c}
\end{eqnarray}
in momentum space.

Next, we consider the following composite operator in flat space:
\begin{eqnarray*}
   \{A^2\} &=& \fr{D-4}{4\b} F_{0\mu\nu} F_0^{\mu\nu} 
        \nonumber \\
           &=& \fr{1}{4} \left[ F_{\mu\nu} F^{\mu\nu} \right] + \fr{{\bar \gm}_2}{2\b} [E_\psi] .
\end{eqnarray*}
This field is related to the trace of energy-momentum tensor as ${\bar \theta}|_{\rm flat} = \b \{A^2\}$, up to the term of gauge-fixing origin which is disregarded because it gives a vanishing contribution in physical correlation functions \cite{freeman}. Note that ${\bar \theta}$ is finite, while $\{A^2\}$ is not so due to the presence of the last term with $1/\b$ in the second line.

The two-point function of $\{A^2\}$ is denoted by $\Gm_{AA}(p^2) = \left\lang \{A^2(p)\} \{A^2(-p)\} \right\rang_{\rm flat}$ in momentum space. Here, although the composite operator $\{A^2\}$ is not finite, the contribution from the term with $1/\b$ vanishes due to the property of $[E_\psi]$. Therefore, $\Gm_{AA}$ is given by the two-point function of the normal product $[F_{\mu\nu}F^{\mu\nu}]$. In such a correlation function, non-local divergences are canceled out and thus it can be written in the form
\begin{eqnarray}
     \Gm_{AA}(p^2) - p^4 \mu^{D-4} \left( \fr{D-4}{\b} \right)^2 L_x = {\rm finite},  \quad 
     L_x = \sum_{n=1}^\infty \fr{x_n(\a)}{(D-4)^n} .
       \label{defining equation of L_x}
\end{eqnarray}
Here, the pure-pole term $L_x$ is defined by this equation. The factor before $L_x$ is introduced for the later convenience. The residue $x_1$ will be directly calculated later.

Since $\b^2 \Gm_{AA}=\lang {\bar \theta} {\bar \theta} \rang_{\rm flat}$, we can see that combining (\ref{theta-theta determining L_c}) and (\ref{defining equation of L_x}), the pure-pole terms satisfy the relation
\begin{eqnarray}
      (D-4)^2 L_x -8 L_c = {\rm finite} .
      \label{relationship between L_x and L_c}
\end{eqnarray}
From this, we obtain the relationship between the residues, 
\begin{eqnarray}
    c_n = \fr{1}{8} x_{n+2}.
      \label{relation between c_n and x_n}
\end{eqnarray} 
This relation means that if the residue $x_3$ is calculated, we can see the residue $c_1$ and then obtain the general $c_n$ from the RG equation (\ref{pole equation of gravity}).

So, we next derive the RG equation that relates $x_3$ with $x_1$. Here, we use the fact that if $F$ is a finite quantity, $\b^{-n} \mu d (\b^n F)/d\mu$ is also finite in spite of the presence of the pole factor $\b^{-n}$ because of $\b^{-n}\mu d\b^n/d\mu = n \a \pd \barb/\pd \a$. Applying this fact to the finite equation (\ref{defining equation of L_x}), we obtain
\begin{eqnarray*}
   \fr{1}{\b^2} \mu \fr{d}{d\mu} \left\{ \b^2 \Gm_{AA}(p^2) - p^4 \mu^{D-4} (D-4)^2 L_x \right\} = {\rm finite} .
\end{eqnarray*}
Since $\b\{A^2\}$ can be described in bare quantities, it satisfies $\mu d(\b \{A^2\})/d\mu = 0$ such that the first term vanishes. Thus, we obtain the RG equation
\begin{eqnarray}
    \fr{1}{\b^2} \mu \fr{d}{d\mu} \left\{ \mu^{D-4} (D-4)^2 L_x \right\} = {\rm finite} .
      \label{RG equation for L_x}
\end{eqnarray}
Expanding this equation and extracting the condition that poles cancel out, we obtain
\begin{eqnarray}
     \fr{\pd}{\pd \a} \left( \a x_2 \right) - \fr{\barb}{\a} \fr{\pd}{\pd \a} \left(\a^2 x_1\right) &=& 0,
          \nonumber \\
    \fr{\pd}{\pd \a} \left( \a x_3 \right) - \fr{\barb}{\a} \fr{\pd}{\pd \a} \left(\a^2 x_2\right) 
        + \fr{\barb^2}{\a^2} \fr{\pd}{\pd \a} \left(\a^3 x_1\right) &=& 0 .
         \label{RG equation of x_2 and x_3}
\end{eqnarray}
Using these equations, we can derive the residues $x_2$ and $x_3$ from $x_1$. As is apparent from the relation (\ref{relation between c_n and x_n}), the equation of $x_n$ for $n \geq 3$ reduces to the same form as (\ref{pole equation of gravity}).

\subsection{Three-point functions}
\noindent

Next, we consider the three-point function of the energy-momentum tensor. Here, we introduce new variable
\begin{eqnarray*}
    {\bar \theta}_2(x,y) = \fr{\dl {\bar \theta}(x)}{\dl \Om(y)} - \half (D-1) \fr{\dl {\bar \theta}(x)}{\dl \chi(y)} , 
\end{eqnarray*}
which satisfies the symmetric condition ${\bar \theta}_2(x,y) = {\bar \theta}_2(y,x)$. In terms of ${\bar \theta}$ and ${\bar \theta}_2$, the condition of the three-point function of $\theta$ (\ref{finiteness condition for theta-theta-theta}) can be written in flat space as
\begin{eqnarray*}
    && \lang {\bar \theta}(x) {\bar \theta}(y) {\bar \theta}(z) \rang_{\rm flat} 
       - \lang {\bar \theta}(x) {\bar \theta}_2(y,z) \rang_{\rm flat}  
       - \lang {\bar \theta}(y) {\bar \theta}_2(z,x) \rang_{\rm flat} 
       - \lang {\bar \theta}(z) {\bar \theta}_2(x,y) \rang_{\rm flat}
             \nonumber \\
    && + \biggl\lang \fr{\dl^3 S}{\dl \Om(x) \dl \Om(y) \dl \Om(z)} \biggr\rang_{\rm flat} = {\rm finite} .
\end{eqnarray*}

The three-point function of $\{A^2\}$ is denoted by $\Gm_{AAA}$. Since ${\bar \theta}|_{\rm flat}=\b\{A^2\}$ and ${\bar \theta}_2(x,y)|_{\rm flat}=-4\b \{A^2\} \dl^D(x-y)$, the condition above can be written in momentum space as
\begin{eqnarray*}
    && \b^3 \Gm_{AAA}(p_x^2,p_y^2,p_z^2) + 4\b^2 \left\{ \Gm_{AA}(p_x^2) + \Gm_{AA}(p_y^2) +\Gm_{AA}(p_z^2) \right\}
            \nonumber \\
    && + b_0 B(p_x,p_y,p_z) + c_0 C(p_x,p_y,p_z) = {\rm finite} .
\end{eqnarray*}
The functions $B$ and $C$ are the contributions from the $G_4$ and $H^2$ terms in the action, respectively, which are defined by
\begin{eqnarray*}
    B(p_x^2,p_y^2,p_z^2) &=& -2(D-2)(D-3)(D-4) 
              \nonumber \\
        && \times     
              \left[ p_x^4 + p_y^4 + p_z^4  -2\left(p_x^2 p_y^2 + p_y^2 p_z^2 + p_z^2 p_x^2 \right) \right] ,
              \nonumber \\     
    C(p_x^2,p_y^2,p_z^2) &=& -4 \left[ (D+2)\left( p_x^4 + p_y^4 + p_z^4 \right) 
                                 +4 \left(p_x^2 p_y^2 + p_y^2 p_z^2 + p_z^2 p_x^2 \right) \right] .
\end{eqnarray*}

In the following, we consider the special cases that some momenta are taken to be on-shell. Combining (\ref{defining equation of L_x}) and (\ref{relationship between L_x and L_c}), we obtain the equations, $\b^3 \Gm_{AAA}(p^2,p^2,0) - 8(D-4)p^4 \mu^{D-4} L_c ={\rm finite}$, and 
\begin{eqnarray}
    && \b^3 \Gm_{AAA}(p^2,0,0) - p^4 \mu^{D-4} \left[ 2(D-2)(D-3)(D-4)L_b + 4(D-6)L_c \right] 
         \nonumber \\
    && = {\rm finite} .
      \label{equations for beta^3 x Gamma_AAA}
\end{eqnarray}

In general, removing the factor $\b^3$, $\Gm_{AAA}$ has the following form:
\begin{eqnarray}
  && \Gm_{AAA}(p_x^2,p_y^2,p_z^2) - \sum {\rm poles} \times \left\{ \Gm_{AA}(p_x^2) + \Gm_{AA}(p_y^2) + \Gm_{AA}(p_z^2) \right\} 
       \nonumber \\
  && - \mu^{D-4} \sum {\rm poles} \times \left\{ \hbox{terms in } p_i^2 p_j^2 \right\} = {\rm finite} .
      \label{general expression of Gm_AAA}
\end{eqnarray}
Since three-point functions with $[E_\psi]$ do not vanish, the term $[E_\psi]/\b$ in $\{A^2\}$ produces non-local poles because of the presence of $1/\b$. Thus, unlike $\Gm_{AA}$, $\Gm_{AAA}$ has non-local poles. The second term in (\ref{general expression of Gm_AAA}) plays an important role to cancel out such non-local poles.

In order to determine the pure-pole factor in front of $\Gm_{AA}$ in (\ref{general expression of Gm_AAA}), we consider the equation obtained by applying $\a \pd/\pd \a$ to (\ref{defining equation of L_x}), which yields the equation for $\Gm_{AAA}(p^2,p^2,0)$ because of $\a \pd S/\pd \a |_{\rm flat} = \int d^D x \{ -\{ A^2 \} + (D-4) [E_A]/2\b -(\pd^\mu A_{0\mu})^2/2\xi_0 \}$ and $\a \pd \{ A^2 \}/\pd \a = -(\a/\b)(\pd \barb/\pd \a) \{A^2\}$. The pole factor can be extracted from this equation and fixed to be $(\a^2/\b) \pd(\barb/\a)/\pd \a$. Therefore, $\Gm_{AAA}(p^2,0,0)$ has the following form:
\begin{eqnarray}
    \Gm_{AAA}(p^2,0,0) - \fr{\a^2}{\b} \fr{\pd}{\pd \a} \left( \fr{\barb}{\a} \right) \Gm_{AA}(p^2)
     - p^4 \mu^{D-4} \left( \fr{D-4}{\b} \right)^3 L_y = {\rm finite} .
     \label{defining equation of L_y}
\end{eqnarray}
Here, the last pure-pole term $L_y$ cannot be deduced from the equation for $\Gm_{AAA}(p^2,p^2,0)$ mentioned above. This term is therefore defined through this equation, which is expanded as
\begin{eqnarray*}
    L_y = \sum_{n=1}^\infty \fr{y_n(\a)}{(D-4)^n} .
\end{eqnarray*}

Multiplying (\ref{defining equation of L_y}) by $\b^3$ and using (\ref{defining equation of L_x}) multiplied by $\b^2$ and (\ref{relationship between L_x and L_c}), we obtain another equation including $\b^3 \Gm_{AAA}$ independent of (\ref{equations for beta^3 x Gamma_AAA}). By eliminating $\b^3 \Gm_{AAA}$ from these equations, we obtain the following pole relation:\footnote{ 
We here correct the typo in \cite{hathrell-QED} on the sign before $2\a^2 \pd (\barb/\a)/\pd \a$ in (\ref{RG equation of L_b}) and the corresponding term in (\ref{defining equation of L_y}). It affects the calculations in Section 5.
} 
\begin{eqnarray}
   && 2(D-2)(D-3)(D-4)L_b + 4 \left[ D-6 -2\a^2 \fr{\pd}{\pd \a} \left( \fr{\barb}{\a} \right) \right] L_c - (D-4)^3 L_y 
             \nonumber \\
   && = {\rm finite} .
       \label{RG equation of L_b}
\end{eqnarray}

Finally, we derive the RG equation for $L_y$. As similar to the derivation of (\ref{RG equation for L_x}), we consider the equation obtained by applying $\b^{-3} \mu d/d\mu$ to (\ref{defining equation of L_y}) multiplied by $\b^3$. Noting that $\mu d(\b^3\Gm_{AAA})/d\mu = \mu d(\b^2\Gm_{AA})/d\mu = 0$, we obtain the following RG equation:
\begin{eqnarray}
   \left( \fr{D-4}{\b} \right)^3  \left[ (D-4)L_y + \b \a \fr{\pd}{\pd \a}L_y \right]
    + \a^2 \fr{\pd^2 \barb}{\pd \a^2} \left( \fr{D-4}{\b} \right)^2 L_x = {\rm finite} ,
        \label{RG equation of L_y}
\end{eqnarray}
where (\ref{defining equation of L_x}) is used.

\section{Reconsiderations of Conformal Anomalies}
\setcounter{equation}{0}
\noindent

Originally, Hathrell considered the three-type of gravitational couplings denoted by $a$, $b$ and $c$, as shown in the previous sections, and he concluded that the $R^2$ divergence appears at $o(\a^3)$ even for QED in curved space.

However, on the other hand, the derived equation (\ref{RG equation of L_b}) gives the relationship between the pure-pole terms $L_b$ and $L_c$ through (\ref{relationship between L_x and L_c}) and (\ref{RG equation of L_y}). So, against his conclusion, his results rather indicate that the independent gravitational counterterms are only two. In this section, we reconsider his results in this context.

We here propose that the gravitational counterterms are given by the two terms as
\begin{eqnarray}
   S_g = \int d^D x \sq{g} \left\{  a_0 F_D + b_0 G_D  \right\} .
    \label{novel gravitational action}
\end{eqnarray}
The novel term $G_D$ is defined by
\begin{eqnarray}
     G_D = G_4 + (D-4) \chi(D) H^2 ,
      \label{definition of G_D}
\end{eqnarray}
where $\chi(D)$ is a finite function of $D$ only and thus this term reduces to the Euler density at $D=4$.

By repeating the previous procedure using the counterterm (\ref{novel gravitational action}) again, we can easily find that the finiteness conditions simply result in the Hathrell's RG equations, (\ref{relationship between L_x and L_c}), (\ref{RG equation for L_x}), (\ref{RG equation of L_b}) and (\ref{RG equation of L_y}), under the relation
\begin{eqnarray}
     L_c - (D-4)\chi(D) L_b = {\rm finite}.
      \label{relation between L_b and L_c}
\end{eqnarray}
The RG equations (\ref{pole equation of gravity}) for $a_n$ and $b_n$ are satisfied, and also for $c_n$ through the relation (\ref{relation between L_b and L_c}), though there is no $\b_c$. Thus, we can make the theory finite using two gravitational counterterms only. In the next section, we will show that the function $\chi$ can be determined completely by solving the coupled RG equations order by order.

On the other hand, we have to pay more attention to the calculation of the finite quantities such as the expression of the conformal anomaly, because the counterterm (\ref{novel gravitational action}) implies that the finite parameter $c$ is eliminated, while extra finite terms are added.

According to the derivation briefly summarized in Appendix A, we find that the expression of the normal product $[F_{\mu\nu}^2]$ in the case of (\ref{novel gravitational action}) can be determined up to the total-divergence term as
\begin{eqnarray}
   \fr{1}{4} [F_{\mu\nu}F^{\mu\nu}] &=& \fr{D-4}{4\b} F_{0\mu\nu}F_0^{\mu\nu} -\fr{{\bar \gm}_2}{2\b} [E_\psi]
        + \fr{D-4}{\b} \mu^{D-4} \Biggl[ \left( L_a + \fr{\barb_a}{D-4} \right) F_D 
         \nonumber \\
   &&   + \left( L_b + \fr{\barb_b}{D-4} \right) G_D   
        - \fr{4\chi(D)(\s +L_\s)}{D-4}\nb^2 H   \Biggr] ,
       \label{intermediate expression of normal product}
\end{eqnarray}
where $\s$ is a finite function of $\a$ and $L_\s=\sum_{n=1}^\infty \s_n/(D-4)^n$, which will be determined below. The factor $\chi$ in the last term is multiplied for convenience.

Using the expression of the normal product (\ref{intermediate expression of normal product}), the trace of the energy-momentum tensor can be expressed in a manifestly finite form as
\begin{eqnarray}
  \theta 
     &=& \fr{D-4}{4} F_{0\mu\nu} F_0^{\mu\nu} + \fr{D-1}{2} E_{0\psi} 
         + (D-4) \left[ a_0 F_D + b_0 \left(G_D -4\chi(D)\nb^2 H \right)\right]
                   \nonumber \\
    &=& \fr{\b}{4} \left[ F_{\mu\nu}F^{\mu\nu} \right] + \half \left( D-1 + {\bar \gm}_2 \right) [E_\psi] 
              - \mu^{D-4} \left( \b_a F_D + \b_b G_D \right) 
                    \nonumber \\
              && -4 \mu^{D-4}\chi(D) \left[ (D-4) b - \s + b_1 \right] \nb^2 H .
          \label{intermediate expression of theta}
\end{eqnarray}
Here, in the second equality, we use the consistency condition to make the last term finite such as
\begin{eqnarray*}
     (D-4)( b+L_b ) - ( \s +L_\s ) = {\rm finite} = (D-4)b-\s +b_1 .
\end{eqnarray*}
The right-hand side just appears in the last term of the expression (\ref{intermediate expression of theta}). The residue $\s_n$ of $L_\s$ is then determined using the residue of $L_b$ as
\begin{eqnarray*}
    \s_n = b_{n+1}   
\end{eqnarray*}
for $n \geq 1$.

Furthermore, in order to determine the finite value $\s$, we consider the finite quantity $\b^{-1} \mu d/d\mu ( \b [ F_{\mu\nu} F^{\mu\nu}]/4)$. Rewriting this expression using the fact that the energy-momentum tensor is independent of $\mu$, we obtain
\begin{eqnarray}
     \fr{1}{\b} \mu \fr{d}{d\mu} \left( \fr{\b}{4} \left[ F_{\mu\nu} F^{\mu\nu} \right] \right)
     &=& - \half \a \fr{\pd {\bar \gm}_2}{\pd \a} [E_\psi] 
           + \mu^{D-4} \left( \a \fr{\pd \barb_a}{\pd \a} F_D 
             + \a \fr{\pd \barb_b}{\pd \a} G_D  \right)
                  \nonumber \\
     &&  + 4 \mu^{D-4} \chi(D) \Biggl\{ \fr{D-4}{\b} \left[ (D-4)b -\s + b_1 \right] 
                  \nonumber \\
     &&  + \fr{1}{\b} \left[ (D-4)\mu \fr{db}{d\mu} -\mu \fr{d\s}{d\mu} + \mu \fr{db_1}{d\mu} \right] 
               \Biggr\} \nb^2 H .
         \label{finiteness condition for sigma}
\end{eqnarray}
From the condition that the last term is finite, we obtain
\begin{eqnarray*}
     \s = \barb_b + b_1
\end{eqnarray*}
and then the inside of the bracket $\{ ~\}$ reduces to the finite value $-\a \pd \barb_b/\pd \a$.

Substituting this result into (\ref{intermediate expression of theta}), we obtain the following simpler expression of conformal anomaly:
\begin{eqnarray}
  \theta = \fr{\b}{4} \left[ F_{\mu\nu} F^{\mu\nu} \right] + \half  \left( D-1+{\bar \gm}_2 \right) \left[ E_\psi \right]
       - \mu^{D-4} \left( \b_a F_D + \b_b E_D \right) ,
        \label{expression of our conformal anomaly}
\end{eqnarray}
where the quantity $E_D$ is defined by
\begin{eqnarray}
   E_D = G_D - 4\chi(D) \nb^2 H .
     \label{definition of E_D}
\end{eqnarray}
The $G_D$ and $\nb^2 H$ terms in the right-hand side of the normal product (\ref{intermediate expression of normal product}) are also unified in this form.

The novel function $E_D$ has a desirable property as the other conformal anomalies $F_D$ and $F_{\mu\nu}^2$ have, which is 
\begin{eqnarray*}
    \fr{\dl}{\dl \Om} \int d^D x \sq{g} E_D =(D-4) E_D .
\end{eqnarray*}
Here, the volume integral of $E_D$ is nothing but the $G_D$ counterterm.

\section{Determination of Gravitational Counterterms}
\setcounter{equation}{0}
\noindent

In this section, we explicitly solve the RG equations and determine the constant $\chi$ order by order.

To determine the pole terms, we need the information of the QED beta function and the simple-pole residues of $L_x$ and $L_y$. In this section, they are expanded as follows:
\begin{eqnarray*}
   \barb &=& \b_1 \a + \b_2 \a^2 + \b_3 \a^3 + o(\a^4),
         \nonumber \\
   x_1 &=& X_1 + X_2 \a + X_3 \a^2 +o(\a^3),
         \nonumber \\
   y_1 &=& Y_1 + Y_2 \a + Y_3 \a^2 +o(\a^3).
\end{eqnarray*}
The specific values of these coefficients will be given in the next section.

We first calculate the residue $x_n$. Using the RG equations (\ref{RG equation of x_2 and x_3}), we can derive $x_2$ and $x_3$ from $x_1$. Furthermore, $x_n$ for $n \geq 4$ can be derived from $x_3$ using the fact that the RG equation of $x_n$ reduces to the same form as (\ref{pole equation of gravity}) for $n \geq 3$ as mentioned before. Using the expressions for $\barb$ and $x_1$ above, we derive the expression of $x_n$ to $o(\a^{n+1})$ for each $n$. For the first several residues, we obtain
\begin{eqnarray}
    x_2 &=& \b_1 X_1 \a + \left( \fr{2}{3}\b_2 X_1 + \b_1 X_2 \right) \a^2 
                + \left( \half \b_3 X_1 + \fr{3}{4} \b_2 X_2 + \b_1 X_3 \right) \a^3 + o(\a^4),
            \nonumber \\
    x_3 &=& -\fr{1}{12} \b_1 \b_2 X_1 \a^3
                + \left( -\fr{1}{15} \b_2^2 X_1  -\fr{1}{10}\b_1 \b_3 X_1 -\fr{1}{20}\b_1 \b_2 X_2 \right) \a^4 + o(\a^5),
            \nonumber \\
    x_4 &=& \fr{1}{20} \b_1^2 \b_2 X_1 \a^4 
                + \left( \fr{31}{360} \b_1 \b_2^2 X_1 + \fr{1}{30} \b_1^2 \b_2 X_2 + \fr{1}{15} \b_1^2 \b_3 X_1 \right) \a^5 
                + o(\a^6) . 
        \label{expression of x_n}
\end{eqnarray}
Note that the lowest term of $x_n$ is given by $o(\a^{n-1})$ for $n \leq 2$, while for $n \geq 3$ it is reduced to $o(\a^n)$. It is probably associated with the fact that the RG equation of $x_n$ becomes simpler for $n \geq 3$. And also, the $o(\a^3)$ term of $x_2$ has the coefficient $X_3$ of 3-loop origin, while the $o(\a^{n+1})$ term of $x_n$ for $n \geq 3$ does not include this coefficient.

The residue $c_n$ is also obtained through the relation $c_n=x_{n+2}/8$ (\ref{relation between c_n and x_n}), and thus $c_1$ starts from $o(\a^3)$.

Next, we calculate $y_n$ to $o(\a^{n+1})$ for each $n$. Expanding the RG equation (\ref{RG equation of L_y}) and evaluating the finiteness condition such that the $n$-th pole term cancels out, we can derive the following relationship between the residues:
\begin{eqnarray}
    && \fr{\pd}{\pd \a} \left( \a y_{n+1} \right) + \barb \a \fr{\pd y_n}{\pd \a}
            \nonumber \\
    &&  + \sum_{m=1}^{n-1} (-1)^m \fr{(m+1)(m+2)}{2} \barb^m \left[ \fr{\pd}{\pd \a} \left( \a y_{n-m+1} \right)
                + \barb \a \fr{\pd y_{n-m}}{\pd \a} \right]
            \nonumber \\
    &&  + (-1)^n \fr{(n+1)(n+2)}{2} \barb^n \fr{\pd}{\pd \a} \left( \a y_1 \right) 
      - \a^2 \fr{\pd^2 \barb}{\pd \a^2} \sum_{m=1}^n (-1)^m m \barb^{m-1} x_{n-m+1} =0 
         \nonumber \\
    &&   \label{pole relation between y_n and x_n} 
\end{eqnarray}
for $n \geq 1$. Since we have already derived the residue $x_n$ from $x_1$, we can derive the residue $y_n$ from $x_1$ and $y_1$ using this equation. The first several residues are given by
\begin{eqnarray}
  y_2 &=& \fr{3}{2} \b_1 Y_1 \a + \left( -\fr{2}{3}\b_2 X_1 +\b_2 Y_1 + \fr{5}{3} \b_1 Y_2 \right) \a^2 
       \nonumber \\
          && + \left( -\fr{3}{2} \b_3 X_1 -\half \b_2 X_2 + \fr{3}{4} \b_3 Y_1 + \fr{5}{4} \b_2 Y_2 
                      + \fr{7}{4} \b_1 Y_3 \right) \a^3 + o(\a^4) ,
       \nonumber \\
  y_3 &=& \half \b_1^2 Y_1 \a^2 
              + \left( -\fr{2}{3} \b_1 \b_2 X_1 + \fr{5}{8} \b_1 \b_2 Y_1 + \fr{2}{3} \b_1^2 Y_2 \right) \a^3
              + \biggl( -\fr{3}{2}\b_1 \b_3 X_1
        \nonumber \\
          &&   -\half \b_1 \b_2 X_2 -\fr{2}{5} \b_2^2 X_1 + \fr{3}{4} \b_1^2 Y_3 
                       + \fr{59}{60} \b_1 \b_2 Y_2 + \fr{1}{5} \b_2^2 Y_1 + \fr{9}{20} \b_1 \b_3 Y_1 
              \biggr) \a^4 
       \nonumber \\
          && + o(\a^5) ,
       \nonumber \\
   y_4 &=& \fr{1}{40} \b_1^2 \b_2 Y_1 \a^4 
               + \biggl( \fr{1}{30} \b_1^2 \b_3 Y_1 + \fr{13}{240} \b_1 \b_2^2 Y_1 + \fr{1}{90} \b_1^2 \b_2 Y_2
                         + \fr{13}{180} \b_1 \b_2^2 X_1 \biggr) \a^5
       \nonumber \\
           && + o(\a^6) ,
       \nonumber \\
   y_5 &=& -\fr{1}{60} \b_1^3 \b_2 Y_1 \a^5 
               + \biggl( -\fr{53}{1260} \b_1^2 \b_2^2 X_1  -\fr{1}{42} \b_1^3 \b_3 Y_1 - \fr{89}{1680} \b_1^2 \b_2^2 Y_1 
       \nonumber \\
           &&   -\fr{1}{126} \b_1^3 \b_2 Y_2 \biggr) \a^6 + o(\a^7) .
         \label{expression of y_n}
\end{eqnarray}
Note that the lowest term of $y_n$ is given by $o(\a^{n-1})$ for $n \leq 3$, while for $n \geq 4$ it starts from $o(\a^n)$. And also, the $o(\a^{n+1})$ term of $y_n$ has the coefficient $Y_3$ of 3-loop origin for $n \leq 3$, while for $n \geq 4$ it does not appear. This result seems to reflect the fact that for $n=k+3$ with $k \geq 1$ the RG equation (\ref{pole relation between y_n and x_n}) reduces to the simpler form 
\begin{eqnarray*}
   \fr{\pd}{\pd \a} \left( \a y_{k+4} \right) + \barb \a \fr{\pd y_{k+3}}{\pd \a} 
   = - \a^2 \fr{\pd^2 \barb}{\pd \a^2} \left( x_{k+3} + \barb x_{k+2} \right) ,
\end{eqnarray*} 
as in the case of $x_n$ for $n \geq 3$.

Now, we can solve the RG equation (\ref{RG equation of L_b}) under the relation (\ref{relation between L_b and L_c}). Expanding (\ref{RG equation of L_b}) and extracting the finiteness condition that poles cancel out, we obtain
\begin{eqnarray}
     4b_{n+1} + 6b_{n+2} + 2b_{n+3} - 8 \left[ 1 + \a^2 \fr{\pd}{\pd \a} \left( \fr{\barb}{\a} \right) \right] c_n
             + 4 c_{n+1} -y_{n+3} = 0
       \label{pole equation for b_n and c_n}
\end{eqnarray}
for $n \geq 1$. Since $y_n$ is related with $c_n$ through (\ref{pole relation between y_n and x_n}) and (\ref{relation between c_n and x_n}), this equation connect $b_n$ with $c_n$. Since the equation for $n \geq 2$ can be derived from the $n=1$ equation using the other RG equations, we use the $n=1$ equation only below.

The $D$-dependent constant $\chi$ is expanded as a power series in $D-4$ such as
\begin{eqnarray*}
   \chi(D) = \sum_{n=1}^\infty \chi_n (D-4)^{n-1} 
           = \chi_1 + \chi_2 (D-4) + \chi_3 (D-4)^2 + \cdots .
\end{eqnarray*}
The relation (\ref{relation between L_b and L_c}) is then expressed as
\begin{eqnarray}
     c_1 &=& \chi_1 b_2 + \chi_2 b_3 + \chi_3 b_4 + \cdots ,
         \nonumber \\
     c_2 &=& \chi_1 b_3 + \chi_2 b_4 + \chi_3 b_5 + \cdots
        \label{linear relation between c_n and b_n}
\end{eqnarray}
and so on. Since $b_n~(n\geq 3)$ can be expressed by $b_2$ using the RG equation (\ref{pole equation of gravity}) for $b_n$, this relation implies that $c_n$ can be obtained from $b_2$.

Since $c_1$ starts from $o(\a^3)$, $b_2$ also starts from $o(\a^3)$. For the moment, $b_2$ is expanded as follows:
\begin{eqnarray*}
    b_2 = B_1 \a^3 + B_2 \a^4 + B_3 \a^5 + o(\a^6) .
\end{eqnarray*}
From the RG equation (\ref{pole equation of gravity}) for $b_n$, we obtain the expressions
\begin{eqnarray*}
    b_3 &=& - \fr{3}{5} \b_1 B_1 \a^4 - \left( \half \b_2 B_1 + \fr{2}{3} \b_1 B_2 \right) \a^5 + o(\a^6) ,
             \nonumber \\
    b_4 &=& \fr{2}{5} \b_1^2 B_1 \a^5 + o(\a^6),  
            \nonumber \\ 
    b_5 &=& o(\a^6)
\end{eqnarray*}
and so on.

Substituting these expressions into the RG equation ({\ref{pole equation for b_n and c_n}) of $n=1$ and expanding up to $\a^5$, we obtain
\begin{eqnarray}
   &&  4 \left( 1 -2\chi_1 \right) B_1 \a^3 + \left\{ 4 \left(1 -2\chi_1 \right) B_2 
                          + \fr{6}{5} \left( -3 - 2\chi_1 + 4 \chi_2 \right) \b_1 B_1 \right\} \a^4
         \nonumber \\
   &&  + \biggl\{ 4 \left(1-2\chi_1 \right) B_3  + \left( -3 - 10 \chi_1 + 4 \chi_2 \right) \b_2 B_1 
          + \fr{4}{3} \left( - 3 - 2 \chi_1 + 4 \chi_2 \right) \b_1 B_2 
         \nonumber \\
   && \qquad           
          + \fr{4}{5} \left( 1 + 2 \chi_2 - 4 \chi_3 \right) \b_1^2 B_1 \biggr\} \a^5 - y_4(\a)  = o(\a^6) .
        \label{RG equation determining chi_n}
\end{eqnarray}
Here, note that the residue $y_4$ starts from $o(\a^4)$. Thus, from the vanishing condition at $o(\a^3)$, the coefficient $\chi_1$ is determined to be 
\begin{eqnarray}
     \chi_1 = \half .
     \label{value of chi_1}
\end{eqnarray}
This is just the result found by Hathrell, which is expressed as $b_2 =2 c_1 + o(\a^4)$ in his paper \cite{hathrell-QED}.

Since $\chi_1=1/2$, the $B_2$-dependence of $o(\a^4)$ in (\ref{RG equation determining chi_n}) disappears. Therefore, substituting the explicit expression of $y_4$ (\ref{expression of y_n}) into (\ref{RG equation determining chi_n}), we obtain the expression
\begin{eqnarray}
    \chi_2 = 1 + \fr{1}{192} \fr{\b_1 \b_2 Y_1}{B_1}
    \label{expression of chi_2}
\end{eqnarray}
from the vanishing condition at $o(\a^4)$. Using the relation (\ref{linear relation between c_n and b_n}) and the result (\ref{value of chi_1}), we can derive
\begin{eqnarray}
   B_1 =  -\fr{1}{48} \b_1 \b_2 X_1 
     \label{expression of B_1}
\end{eqnarray}
from the expression of $c_1$ at $o(\a^3)$ which can be read from (\ref{expression of x_n}) through the relation (\ref{relation between c_n and x_n}). Substituting this expression, we obtain
\begin{eqnarray}
    \chi_2 = 1 -\fr{Y_1}{4X_1} .
    \label{expression of chi_2}
\end{eqnarray}

Using the result (\ref{expression of chi_2}), the coefficient $B_2$ can be calculated from the expression of $c_1$ at $o(\a^4)$ as
\begin{eqnarray}
   B_2 = - \fr{1}{160} \left( 4 \b_1^2 \b_2 X_1 + \fr{8}{3} \b_2^2 X_1 + 4 \b_1 \b_3 X_1 + 2 \b_1 \b_2 X_2  
                                 - \b_1^2 \b_2 Y_1 \right) .
          \label{expression of B_2}
\end{eqnarray}

Furthermore, since the $B_3$-dependence in (\ref{RG equation determining chi_n}) disappears due to (\ref{value of chi_1}), we can solve the condition (\ref{RG equation determining chi_n}) at $o(\a^5)$ using the expressions of (\ref{expression of B_1}), (\ref{expression of B_2}) and (\ref{expression of chi_2}). Thus, we obtain
\begin{eqnarray}
   \chi_3 = \fr{1}{8} \left( 2- \fr{Y_1}{X_1} \right) \left( 3 -\fr{Y_1}{X_1} \right) 
             - \fr{1}{6} \fr{\b_2}{\b_1^2} \left( 1 - \fr{Y_1}{X_1} \right) 
             + \fr{1}{6} \fr{X_2}{\b_1 X_1} \left( \fr{Y_2}{X_2} - \fr{3}{2} \fr{Y_1}{X_1} \right) .
       \label{expression of chi_3}
\end{eqnarray}
As a result, $\chi_3$ does not depend on $\b_3$.

In this way, we can determine the coefficient $\chi_n$ order by order.

\section{Values of The Parameters}
\setcounter{equation}{0}
\noindent

Let us determine the coefficients $\chi_n$ and the residues of pole terms by substituting the concrete values. The coefficients of the beta function up to 3-loop order are computed as \cite{gkls}
\begin{eqnarray}
    \b_1 = \fr{8}{3} \fr{1}{4\pi}, \qquad \b_2 = 8 \fr{1}{(4\pi)^2}, \qquad \b_3 = -\fr{124}{9}\fr{1}{(4\pi)^3}.
     \label{QED beta function parameter}
\end{eqnarray}
The first two are used to determine $\chi_{1,2,3}$ below and the last is necessary to calculate the residue $b_n$ to $o(\a^{n+2})$ for each $n$.

The values of $X_{1,2}$ and $Y_{1,2}$ are obtained from the direct 2-loop computations of $\Gm_{AA}$ and $\Gm_{AAA}$, respectively \cite{hathrell-QED}. The function $\Gm_{AA}$ is calculated as
\begin{eqnarray*}
   && \Gm_{AA}(p^2) (2\pi)^D \dl^D(p+q) 
           \nonumber \\
   && = \left( \fr{D-4}{\b} Z_3 \right)^2 \fr{1}{4} \int \fr{d^D k}{(2\pi)^D} \fr{d^D l}{(2\pi)^D}
                         K^{\mu\nu}(k,k-p) K^{\lam\s}(l,l-q) 
            \nonumber \\
   && \qquad\qquad\qquad\qquad\qquad  \times
         \lang A_\mu (k) A_\nu (p-k) A_\lam (l) A_\s (q-l) \rang_{\rm flat} ,
\end{eqnarray*}
where 
\begin{eqnarray*}
   K^{\mu\nu}(k,k-p)=k \cdot (k-p) \dl^{\mu\nu} -(k-p)^\mu k^\nu .
\end{eqnarray*}
The renormalization factor $Z_3$ arises by replacing $F_{0\mu\nu}$ in $\{A^2\}$ with $Z_3^{1/2}F_{\mu\nu}$. The four-point function of $A_\mu$ is evaluated up to $o(\a)$ for the diagrams such that two composite operators are connected. Carrying out the momentum integrals, we obtain 
\begin{eqnarray*}
    \Gm_{AA}(p^2) = \fr{p^4 \mu^{D-4} }{(4\pi)^2} \left\{ - \half \fr{1}{D-4} + \fr{\a}{4\pi} \left( \fr{4}{3} \fr{1}{(D-4)^2} 
             + \fr{5}{3} \fr{1}{D-4} \right) \right\} + {\rm finite} .
\end{eqnarray*}
From this expression, we obtain the coefficients of $x_1$ as
\begin{eqnarray}
     X_1 = - \half \fr{1}{(4\pi)^2}, \qquad X_2 = \fr{5}{3} \fr{1}{(4\pi)^3} .
         \label{X_1 and X_2}
\end{eqnarray}
Taking account of the factor $((D-4)/\b)^2$ in (\ref{defining equation of L_x}) introduced for convenience, the lowest order term of $x_2$ is also determined to be $-4\a/3(4\pi)^3$, which is consistent with the RG equation (\ref{RG equation of x_2 and x_3}).

Similarly, the three-point function $\Gm_{AAA}$ with two on-shell momenta is calculated as
\begin{eqnarray*}
   \Gm_{AAA}(p^2,0,0) = \fr{p^4 \mu^{D-4}}{(4\pi)^2} \left\{ - \half \fr{1}{D-4} + \fr{\a}{4\pi} \left( 2 \fr{1}{(D-4)^2} 
             + \fr{11}{6} \fr{1}{D-4} \right) \right\} + {\rm finite}.
\end{eqnarray*}
From this expression, we obtain the coefficients of $y_1$ as
\begin{eqnarray}
     Y_1 = - \half \fr{1}{(4\pi)^2}, \qquad Y_2 = \fr{11}{6} \fr{1}{(4\pi)^3} .
      \label{Y_1 and Y_2}
\end{eqnarray}
The lowest order term of $y_2$ is determined to be $-2\a/(4\pi)^3$ from (\ref{defining equation of L_y}), which is consistent with the RG equation (\ref{pole relation between y_n and x_n}).

Substituting the values of $X_{1,2}$ and $Y_{1,2}$ into (\ref{expression of chi_2}) and (\ref{expression of chi_3}), we finally obtain
\begin{eqnarray}
     \chi_2 = \fr{3}{4},  \qquad \chi_3 = \fr{1}{3} .
     \label{result of chi(D)}
\end{eqnarray}

Now, we give some comments on the universality of the function $\chi$. First, the value $\chi_1=1/2$ is probably independent of the theory. It has been confirmed for conformally coupled massless scalar theory \cite{hathrell-scalar} and Yang-Mills theory \cite{freeman}. Especially for Yang-Mills theory, the other coefficients of $\chi$ may also be the same as (\ref{result of chi(D)}) because the residues of pole terms satisfy almost the same RG equations as those of QED.\footnote{ 
On the other hand, as for scalar theory, we are afraid that some uncertainty in the coupling with gravity may be left yet.
} 

Furthermore, $\chi_1$ and $\chi_2$ agree with those conjectured in the model of quantum gravity \cite{hamada02}, but $\chi_3$ unfortunately disagrees. It seems that the condition imposed to determine the action $G_D$ in \cite{hamada02} may be somewhat strong. However, the difference is of higher orders and does not affect the loop calculations done there. Thus, the result is also consistent with quantum corrections including gravity.

\section{Gravitational Effective Action}
\setcounter{equation}{0}
\noindent

Finally, we discuss the properties of the conformal anomaly $E_D$ and its physical implications to the effective action.

Consider the conformal variation $\dl_\om g_{\mu\nu} = 2\om g_{\mu\nu}$ of the gravitational effective action $\Gm$ as
\begin{eqnarray*}
   \dl_\om \Gm = \int d^D x \sq{g} \om 
                 \left\{ \eta_1 R_{\mu\nu\lam\s}^2 + \eta_2 R_{\mu\nu}^2  + \eta_3 R^2 + \eta_4 \nb^2 R \right\} .
\end{eqnarray*}
The right-hand side describes possible expressions of conformal anomalies. The Wess-Zumino consistency condition \cite{wz, bcr} in $D$ dimensions, $[ \dl_{\om_1}, \dl_{\om_2} ] \Gm =0$, gives the condition for the parameters as \cite{hamada02}
\begin{eqnarray*}
    4 \eta_1 + D \eta_2 + 4(D-1) \eta_3 + (D-4) \eta_4 = 0 .
\end{eqnarray*}
Three independent combinations satisfying this equation are given by the square of the Weyl tensor in $D$ dimensions $F_D$, the usual Euler density $G_4$ and 
\begin{eqnarray*}
   M_D = (D-4) H^2 - 4 \nb^2 H .
\end{eqnarray*}
Note that $M_D$ corresponds to what is called the trivial conformal anomaly, but it is no longer trivial in $D$ dimensions. The function $E_D$ can be written in a linear combination of the usual Euler density and this function as $E_D = G_4 + \chi(D) M_D$.

Let us consider the four-dimensional limit in the following. Using the value (\ref{value of chi_1}), we find that the function $E_D$ reduces to the form
\begin{eqnarray}
    E_4 = G_4 - \fr{2}{3} \nb^2 R .
       \label{expression of E_4}
\end{eqnarray}
This is just the combination proposed by Riegert \cite{riegert}. When the metric field is decomposed into the conformal factor and others as $g_{\mu\nu}=e^{2\phi}\bg_{\mu\nu}$, the function (\ref{expression of E_4}) satisfies the relation $\sq{g}E_4 = \sq{\bg}(4\bDelta_4 \phi + \bE_4)$, where $\sq{g}\Delta_4$ is a conformally invariant fourth-order differential operator for a scalar quantity defined by
\begin{eqnarray*}
   \Delta_4 = \nb^4 + 2 R^{\mu\nu} \nb_\mu \nb_\nu -\fr{2}{3}R \nb^2 + \fr{1}{3} \nb^\mu R \nb_\mu .
\end{eqnarray*}
The non-local action obtained by integrating the conformal anomaly $b_1 \sq{g}E_4$ over the conformal mode $\phi$ is expressed as \cite{riegert}
\begin{eqnarray}
    \fr{b_1}{8} \int d^4 x \sq{g} E_4 \fr{1}{\Delta_4} E_4 .
        \label{Riegert non-local action}
\end{eqnarray}
This action is the four-dimensional version of the Polyakov's non-local action $\int d^2 x \sq{g} R \Delta_2^{-1}R$ \cite{polyakov}, where $\Delta_2=-\nb^2$ is a conformally invariant operator in two dimensions.

The local part of (\ref{Riegert non-local action}) is given by $b_1 \sq{\bg} ( 2\phi \bDelta_4 \phi + {\bar E}_4 \phi )$, called the Riegert action. Thus, the kinetic term of the conformal mode is induced quantum mechanically. As similar to two dimensional gravity, the Riegert action can be quantized \cite{am,amm92,amm97,hh,hamada12,hamada12RxS3} and it has been known that the combined system of the Riegert and the Weyl actions generates the BRST operator of quantum diffeomorphism that imposes for physical field operators to be, in CFT terminology, Hermitian primary scalars only \cite{hamada12,hamada12RxS3}.

\section{Conclusions}
\setcounter{equation}{0}
\noindent

One of the significant observations that should be emphasized here is as follows. Classically, there is some uncertainty in how to choose the combinations of the fourth-order gravitational actions and their dimensionless coupling constants. When going to quantum field theory, however, it is possible to settle the problem of uncertainty by imposing the finiteness condition of renormalization.

In this paper, reconsidering the Hathrell's RG equations, we determined the expressions of the gravitational counterterms (\ref{novel gravitational action}) and the conformal anomalies (\ref{expression of our conformal anomaly}) for the dimensionally regularized QED in curved spacetime. We showed that at all orders of the perturbation, the independent expressions of them are only two: the square of the Weyl tensor in $D$ dimensions $F_D$ (\ref{square of Weyl tensor in D dimensions}) and the modified Euler density $E_D$ (\ref{definition of E_D}) whose bulk part is given by $G_D$ (\ref{definition of G_D}). The $D$-dependent constant $\chi(D)$ can be determined order by order in series of $D-4$, whose first several terms were calculated explicitly as
\begin{eqnarray*}
    \chi(D) = \half + \fr{3}{4}(D-4) + \fr{1}{3} (D-4)^2 + o((D-4)^3) .
\end{eqnarray*}

The number of the gravitational coupling constants was reduced to two. The situation will be maintained in conformally coupled theories. This is one of the results we have hoped for, because we think that the number of the couplings is too many to describe the dynamics of gravity.
Especially, the elimination of the coupling $c$ before the $R^2$ action is significant to describe the dynamics of the conformal factor which will govern the evolution of the early universe.

Unlike the Hathrell's result (\ref{expression of Hathrell conformal anomaly}), the final expression of conformal anomaly (\ref{expression of our conformal anomaly}) has a suitable structure that when taking the $D \to 4$ limit, the dependence on the unspecified parameters $\mu$, $a$ and $b$ all disappears and the ambiguous $\nb^2 R$ term is fixed in the form $E_4$ (\ref{expression of E_4}). Since $\chi(4)=1/2$ is a constant independent of the theory, the combination $E_4$ is probably a universal expression of conformal anomaly at four dimensions.

Finally, we summarize the residues for the counterterm $G_D$, which were determined using the QED beta function up to 3 loop order as 
\begin{eqnarray*}
    b_1 &=& \fr{73}{360} \fr{1}{(4\pi)^2} - \fr{1}{6} \fr{\a^2}{(4\pi)^4} + \fr{25}{108} \fr{\a^3}{(4\pi)^5} + o(\a^4),
       \nonumber \\
    b_2 &=& \fr{2}{9} \fr{\a^3}{(4\pi)^5} + \fr{22}{135} \fr{\a^4}{(4\pi)^6} + o(\a^5) .
\end{eqnarray*} 
Here, $b_2$ is obtained from the expressions (\ref{expression of B_1}) and (\ref{expression of B_2}) and $b_1$ is calculated from $b_2$ using the RG equation (\ref{pole equation of gravity}) for $b_n$. The constant independent of $\a$ in $b_1$ cannot be determined from the RG equation, which is calculated from the direct 1-loop calculation. The $o(\a^3)$ term of $b_1$ and the $o(\a^4)$ term of $b_2$ are new results. The other residues are summarized in Appendix D. For the completion, we also add the value of the residue $a_1$ \cite{dh},
\begin{eqnarray*}
     a_1 = -\fr{3}{20} \fr{1}{(4\pi)^2} - \fr{7}{72} \fr{\a}{(4\pi)^3} + o(\a^2) , 
\end{eqnarray*}
which can be calculated using (\ref{finiteness condition for theta^munu theta_munu}) in Appendix C.



\appendix

\section{Derivation of The Normal Product $[F_{\mu\nu}F^{\mu\nu}]$}
\setcounter{equation}{0}
\noindent

We here briefly summarize how to derive the expression of the normal product $[F_{\mu\nu}F^{\mu\nu}]$ \cite{hathrell-QED}.

First, we consider the finite quantity obtained by applying $\xi \pd/\pd\xi$ to the renormalized correlation function $\lang \prod^{N_A}_{j=1} A_{\mu_j}(x_j) \prod^{N_\psi}_{k=1} ( \psi ~{\rm or}~{\bar \psi} )(x_k) \rang$. We then obtain
\begin{eqnarray}
   && \left\lang \int d^D x \sq{g} \left\{ \fr{1}{\xi} \left( \nb^\mu A_\mu \right)^2  -[E_\psi] \xi \fr{\pd}{\pd \xi} \log Z_2  \right\}
        \prod^{N_A}_{j=1} A_{\mu_j}(x_j)  \prod^{N_\psi}_{k=1} \left( \psi ~{\rm or}~{\bar \psi} \right)(x_k) \right\rang 
          \nonumber \\
   && = {\rm finite} .
       \label{finite combination with gauge parameter}
\end{eqnarray}

Next, in order to obtain a finite expression including $F_{0\mu\nu}F^{\mu\nu}_0$, we consider the finite quantity derived by applying $\a \pd/\pd\a$ to the renormalized correlation function. The $\a$-dependences of various bare parameters are calculated as $\a \pd e_0/\pd \a = (D-4)e_0/2\b$, $\a \pd \xi_0/\pd \a = \xi_0 \barb/\b$, $\a \pd (\log Z_3)/\pd \a = \barb/\b$ and $\a \pd (\log Z_2)/\pd \a = [ \gm_2 + \barb \xi \pd(\log Z_2)/\pd \xi ]/\b$ for the QED sector. For the gravity sector, we obtain $\a \pd a_0/\pd \a = - \mu^{D-4} [(D-4) L_a + \barb_a ]/\b$ and similar equations for $b_0$ and $c_0$. Using these, we finally obtain
\begin{eqnarray*}
   && \Biggl\lang \int d^D x \sq{g}  \Biggl\{ \fr{D-4}{4\b} F_{0\mu\nu}F_0^{\mu\nu} -\fr{{\bar \gm}_2}{2\b} E_{0\psi}
         + \fr{D-4}{\b} \mu^{D-4} \Biggl[ \left( L_a + \fr{\barb_a}{D-4} \right) F_D 
          \nonumber \\
   &&    + \left( L_b + \fr{\barb_b}{D-4} \right) G_4  + \left( L_c + \fr{\barb_c}{D-4} \right) H^2 \Biggr] \Biggr\} 
        \prod^{N_A}_{j=1} A_{\mu_j}(x_j) \prod^{N_\psi}_{k=1} \left( \psi ~{\rm or}~{\bar \psi} \right)(x_k) \Biggr\rang
           \nonumber \\
   &&  = {\rm finite} .
\end{eqnarray*}
Here, we use the fact that $N_A$ and $N_\psi$ can be replaced with the volume integrals of the equation-of-motion operators $E_{0A}$ and $E_{0\psi}$ (\ref{equation-of-motion operator}) in the correlation function. The interaction term $e_0 {\bar \psi}_0 \gm^\mu \psi_0 A_{0\mu}$ is also eliminated by using $E_{0A}$ and then the kinetic term of gauge field appears. The finite combination (\ref{finite combination with gauge parameter}) and the apparently finite quantity put away to the right-hand side.

This equation means that the inside of the bracket $\{~\}$ is the normal ordered quantity up to total divergences. Here, noting that $(D-4)/\b =1+\sum {\rm poles}$ (\ref{pole expansion of 1/beta}), it has the structure of the normal product mentioned in the text and thus it is identified with $[F_{\mu\nu}F^{\mu\nu}]/4$. Since the candidate for the total divergence term is only $\nb^2 H$, we obtain
\begin{eqnarray}
   \fr{1}{4} [F_{\mu\nu}F^{\mu\nu}] &=& \fr{D-4}{4\b} F_{0\mu\nu}F_0^{\mu\nu} -\fr{{\bar \gm}_2}{2\b} E_{0\psi}
        + \fr{D-4}{\b} \mu^{D-4} \Biggl[ \left( L_a + \fr{\barb_a}{D-4} \right) F_D 
         \nonumber \\
   &&   + \left( L_b + \fr{\barb_b}{D-4} \right) G_4 + \left( L_c + \fr{\barb_c}{D-4} \right) H^2  
        - \fr{4(\s +L_\s)}{D-4}\nb^2 H   \Biggr] .
         \nonumber \\
   &&   \label{normal product of F^2}
\end{eqnarray}
Here, $\s$ is a finite function of $\a$ and $L_\s$ is the pure-pole term, which are defined through this equation. These quantities are determined by imposing other finiteness conditions. The results are given in the text.

\section{The Equation-of-Motion Operators}
\setcounter{equation}{0}
\noindent

The equation-of-motion operators for gauge and fermion fields are defined, respectively, by
\begin{eqnarray}
    E_{0A} &=& \fr{1}{\sq{g}} A_{0\mu} \fr{\dl S}{\dl A_{0\mu}}
           = A_{0\mu} \nb_\nu F_0^{\mu\nu} -e_0 {\bar \psi_0} \gm^\mu A_{0\mu} \psi_0 
              - \fr{1}{\xi_0} A_{0\mu} \nb^\mu \nb^\nu A_{0\nu} ,
                 \nonumber \\
    E_{0\psi} &=& \fr{\dl S}{\dl \chi}
           \equiv \fr{1}{\sq{g}} \left( {\bar \psi}_0 \fr{\dl S}{\dl {\bar \psi}_0} + \psi_0 \fr{\dl S}{\dl \psi_0} \right) 
           = 2i {\bar \psi}_0 \!\! \stackrel{\leftrightarrow}{D\!\!\!\!/} \psi_0 ,
         \label{equation-of-motion operator}
\end{eqnarray}
where covariant derivative with arrow $\stackrel{\leftrightarrow}{D_\mu}$ is defined by replacing $\pd_\mu$ in $D_\mu$ with $( \overrightarrow{\pd_\mu} - \overleftarrow{\pd_\mu} )/2$.

Although the equation-of-motion operators are written in terms of the bare fields, they are finite in correlation functions. It is demonstrated in the path integral formalism as follows. Carrying out an integration-by-part, we obtain the following relations: 
\begin{eqnarray}
    \left\lang E_{0A}(x) \prod^{N_A}_{j=1} A_{\mu_j} (x_j) \right\rang 
    &=& \sum_{j=1}^{N_A} \fr{1}{\sq{g}} \dl^D (x-x_j) \left\lang \prod^{N_A}_{j=1} A_{\mu_j} (x_j) \right\rang ,
             \nonumber \\
        \left\lang E_{0\psi}(x) \prod^{N_\psi}_{j=1} \left( \psi ~{\rm or}~ {\bar \psi} \right) (x_j) \right\rang
    &=& \sum^{N_\psi}_{j=1} \fr{1}{\sq{g}} \dl^D (x-x_j) 
         \left\lang \prod^{N_\psi}_{j=1} \left( \psi ~{\rm or}~ {\bar \psi} \right) (x_j) \right\rang  .           
             \nonumber \\
    &&     \label{correlator with EOM operator}
\end{eqnarray}
Here, note that there is no term from functional differentials at the same point, because it is dimensionally regularized to zero such as $\dl A_\mu(x)/\dl A_\nu(x)=\dl^\mu_{~\nu}\dl^D(0)=0$. The right-hand sides are obviously finite and thus the left-hand sides are also finite. So, the equation-of-motion operators can be written in terms of the normal products as 
\begin{eqnarray*}
    E_{0A} = [E_A], \qquad E_{0\psi} = [E_\psi] .
\end{eqnarray*}
From (\ref{correlator with EOM operator}), $\int d^D x \sq{g} E_{0A}$ and $\int d^D x \sq{g} E_{0\psi}$ can be replaced with the numbers $N_A$ and $N_\psi$, respectively, in correlation functions.

\section{Finiteness Conditions for Two and Three-Point Functions}
\setcounter{equation}{0}
\noindent

The energy-momentum tensor is defined by $\theta^{\mu\nu} = (2/\sq{g}) \dl S/\dl g_{\mu\nu}$ and its trace is denoted by $\theta= \theta^{\mu}_{~\mu}=\dl S/\dl \Om$. The energy-momentum tensor of the QED sector is given by 
\begin{eqnarray*}
   \theta_{\rm QED}^{\mu\nu}
       = - F_0^{\mu\lam}F_{0 \lam}^\nu + \fr{1}{4} g^{\mu\nu} F_{0 \lam\s}F_0^{\lam\s}
            - \fr{i}{2} {\bar \psi}_0 \left( \gm^\mu \! \! \stackrel{\leftrightarrow}{D^\nu} 
                              + \gm^\nu \! \! \stackrel{\leftrightarrow}{D^\mu} 
                              -2 g^{\mu\nu} \!\! \stackrel{\leftrightarrow}{D\!\!\!\!/}  \right) \psi_0 
\end{eqnarray*}
and its trace is 
\begin{eqnarray*}
    \theta_{\rm QED} = (D-4) \fr{1}{4} F_{0\mu\nu} F_0^{\mu\nu}
              + (D-1) i {\bar \psi}_0 \!\! \stackrel{\leftrightarrow}{D\!\!\!\!/} \psi_0 .
\end{eqnarray*}
Here, we disregard the term of gauge-fixing origin because it gives no contribution in physical correlation functions.

Since the partition function is finite, its gravitational variations are also finite. Thus, carrying out the variation two times, we obtain 
\begin{eqnarray*}
    \lang \theta^{\mu\nu}(x) \theta^{\lam\s}(y) \rang 
    - \fr{2}{\sq{g(y)}} \biggl\lang \fr{\dl \theta^{\mu\nu}(x)}{\dl g_{\lam\s}(y)} \biggr\rang
    = {\rm finite} .
\end{eqnarray*}

Taking the flat space limit and going to momentum space, we obtain the following condition: 
\begin{eqnarray*}
   \lang \theta^{\mu\nu}(p) \theta^{\lam\s}(-p) \rang_{\rm flat} - a_0 A^{\mu\nu,\lam\s}(p) -c_0 C^{\mu\nu,\lam\s}(p) = {\rm finite} ,
\end{eqnarray*}
where the functions $A^{\mu\nu,\lam\s}$ and $C^{\mu\nu,\lam\s}$ are defined by
\begin{eqnarray*}
   A^{\mu\nu,\lam\s}(p) 
      &=& \fr{4(D-3)}{D-2}\Bigl[ p^4 \bigl(\dl^{\mu\lam}\dl^{\nu\s}+\dl^{\mu\s}\dl^{\nu\lam} \bigr) 
          -p^2 \bigl( \dl^{\mu\lam} p^\nu p^\s + \dl^{\mu\s} p^\nu p^\lam + \dl^{\nu\lam} p^\mu p^\s 
                 \nonumber \\
      &&  +\dl^{\nu\s} p^\mu p^\lam \bigr)  + 2 p^\mu p^\nu p^\lam p^\s  \Bigr]
          - \fr{8(D-3)}{(D-1)(D-2)} \Bigl[ p^4 \dl^{\mu\nu}\dl^{\lam\s} 
                \nonumber \\
      && - p^2 \bigl( \dl^{\mu\nu} p^\lam p^\s + \dl^{\lam\s} p^\mu p^\nu \bigr) + p^\mu p^\nu p^\lam p^\s \Bigr] ,
             \nonumber \\
   C^{\mu\nu,\lam\s}(p) 
      &=& \fr{8}{(D-1)^2} \Bigl[ p^4 \dl^{\mu\nu}\dl^{\lam\s} 
        - p^2 \bigl( \dl^{\mu\nu} p^\lam p^\s + \dl^{\lam\s} p^\mu p^\nu \bigr) + p^\mu p^\nu p^\lam p^\s \Bigr] ,
\end{eqnarray*}
which are derived from the $F_D$ and $H^2$ terms in the action, respectively. Contracting the indices of the energy-momentum tensor, we obtain
\begin{eqnarray}
   \lang \theta(p) \theta(-p) \rang_{\rm flat} - 8 c_0 p^4 = {\rm finite}
      \label{finiteness condition for theta-theta}
\end{eqnarray}
and
\begin{eqnarray}
   \lang \theta^{\mu\nu}(p) \theta_{\mu\nu}(-p) \rang_{\rm flat} - 4(D-3)(D+1)a_0 p^4 -\fr{8}{D-1} c_0 p^4 = {\rm finite} .
     \label{finiteness condition for theta^munu theta_munu}
\end{eqnarray}

And also from the variation of the partition function with respect to $\Om$ three times, we obtain
\begin{eqnarray}
    && \lang \theta(x) \theta(y) \theta(z) \rang - \biggl\lang \fr{\dl \theta(x)}{\dl \Om(y)} \theta(z) \biggr\rang
       - \biggr\lang \fr{\dl \theta(y)}{\dl \Om(z)} \theta(x) \biggr\rang 
       - \biggl\lang \fr{\dl \theta(z)}{\dl \Om(x)} \theta(y) \biggr\rang
             \nonumber \\
    && + \biggl\lang \fr{\dl \theta(x)}{\dl \Om(y) \dl \Om(z)} \biggr\rang = {\rm finite} .
      \label{finiteness condition for theta-theta-theta}
\end{eqnarray}

\section{Values of The Residues}
\setcounter{equation}{0}
\noindent

Substituting the values (\ref{QED beta function parameter}) and (\ref{X_1 and X_2}) into the expression of $x_n$ (\ref{expression of x_n}), we obtain
\begin{eqnarray*}
   x_1(\a) &=&  - \fr{1}{2} \fr{1}{(4\pi)^2} + \fr{5}{3} \fr{\a}{(4\pi)^3} + o(\a^2) ,
         \nonumber \\
   x_2(\a) &=& - \fr{4}{3} \fr{\a}{(4\pi)^3} + \fr{16}{9} \fr{\a^2}{(4\pi)^4} + o(\a^3) ,
         \nonumber \\
   x_3(\a) &=& \fr{8}{9} \fr{\a^3}{(4\pi)^5} - \fr{40}{27} \fr{\a^4}{(4\pi)^6} + o(\a^5) , 
         \nonumber \\
   x_4(\a) &=& -\fr{64}{45} \fr{\a^4}{(4\pi)^6} - \fr{224}{243} \fr{\a^5}{(4\pi)^7} + o(\a^6) .
\end{eqnarray*}

Substituting (\ref{QED beta function parameter}), (\ref{X_1 and X_2}) and (\ref{Y_1 and Y_2}) into the expression of $y_n$ (\ref{expression of y_n}), we obtain
\begin{eqnarray*}
   y_1(\a) &=&  - \fr{1}{2} \fr{1}{(4\pi)^2} + \fr{11}{6} \fr{\a}{(4\pi)^3} + o(\a^2) ,
         \nonumber \\
   y_2(\a) &=& - 2 \fr{\a}{(4\pi)^3} + \fr{184}{27} \fr{\a^2}{(4\pi)^4} + o(\a^3) ,
         \nonumber \\
   y_3(\a) &=& -\fr{16}{9} \fr{\a^2}{(4\pi)^4} + \fr{740}{81} \fr{\a^3}{(4\pi)^5} + o(\a^4) , 
         \nonumber \\
   y_4(\a) &=& -\fr{32}{45} \fr{\a^4}{(4\pi)^6} - \fr{9712}{1215} \fr{\a^5}{(4\pi)^7} + o(\a^6) ,
         \nonumber \\
   y_5(\a) &=& \fr{512}{405} \fr{\a^5}{(4\pi)^7} + \fr{416128}{25515} \fr{\a^6}{(4\pi)^8} + o(\a^7) .
\end{eqnarray*}


\end{document}